\documentclass[12pt, twocolumn]{article}
\usepackage{hyperref}
\usepackage[margin=1in]{geometry}
\usepackage{graphicx}
\usepackage{subcaption}
\usepackage{float}
\usepackage{multicol}
\begin{document}

\twocolumn[
  \begin{@twocolumnfalse}

\author{Justin Downes \\ jdownes4@gmu.edu}

\title{Evolving Networks Created by Preferential Attachment and Decay}

\date{}

\maketitle
\begin{abstract}
Growing synthetic networks that follow power law distributions of a node's degree often involves adding one node at a time. Each node is added to the network with a fixed amount of edges and those edges are frozen for all future time steps. Yet real world networks often continuously evolve with edges being added and removed while new nodes are added to the network. Many existing growth models based on preferential attachment do not account for this evolutionary capability and when you extend their growth methods to add and remove edges to existing nodes the node degree distribution quickly loses its scale-free structure. This paper will go over a method to extend well known preferential attachment growth models to allow for the evolution of edges within a network while still maintaining a power law node degree distribution. \\ \\
\end{abstract}
 \end{@twocolumnfalse}
]

\section{Introduction}

Many growth models have been proposed to explain the formation of a variety of structures found in real world networks. These real networks often display a scale free structure, where the distribution of node degrees follows a power-law. Often these models grow networks by adding nodes one at a time with a variable number of edges introduced from the new node to existing nodes. This growth structure makes sense for situations such as citation networks where created edges do not change over time, but fail to adequately describe the evolution of many social networks where edges are created and removed constantly even while new nodes are added. 

When it comes to social networks, having tools that can successfully mimic their structure has some unique benefits not found in other domains. To understand social network dynamics, scientists want to work with real data, but with that real data comes important concerns and complications. User privacy and data access prevent much of our ability to capture structure on popular networks \cite{onlineSocial}. And data granularity and missing data can force the implementation of small models based on sub groups within the network \cite{onlineSocial}. Simulating this network data can provide the ability to develop and refine models before having access to real data, which once having gained access too, will help refine the simulation models themselves. Additionally, while waiting for data to be collected, models can be trained and developed on simulated networks and then evaluated on real world network data \cite{socialSim}.

In this study I seek to demonstrate how existing growth models can be used in situations where a network is evolved, that is existing nodes create and lose edges at each time step. The specific growth models I am looking at are ones that are based off of preferential attachment. This paper describes an evolutionary model that works in conjunction with the preferential attachment growth model. The goal of this evolution is that at any given time step of evolution the degree distribution follows closely to the distribution of the network that was originally grown by one of the base preferential attachment growth models. Additionally, I will show what the implications are when using these evolutionary models with regards to the power law distribution of node degrees, that is how they diverge from the original growth model's distribution. Finally I will attempt to show what modifications to a model are sufficient to evolve a network that replicates the structure of one grown through the preferential attachment growth models. 

We know that many real networks follow a scale-free structure and that those networks do not necessarily have static edges\cite{prefDeletion}. By developing models that can simulate this situation we can come closer to simulating real networks and augment the incredibly difficult task of collecting real world social network data.This will help increase the pace at which computational social science models are developed and provide more data sets to evaluate those models. 

\section{Background}
One method to grow networks that follow scale-free structures is for nodes to follow preferential attachment when being added to the network and connected to existing nodes. Preferential attachment is the concept that the likelihood of an edge being added to a node increases with the degree of that node \cite{statMech}. Therefore, when a node is added to a network it is more likely to be connected to nodes that already have a lot of connections. This study will cover three such growth models and how they can be used to evolve existing networks. Evolving networks refers to the creation and removal of edges between existing nodes, which is sometimes referred to as dynamic networks. Whereas growing networks refers to adding new nodes and connecting them to existing nodes. The models used in this study, the Barab\'asi-Albert model \cite{statMech}, the Bianconi-Barab\'asi model \cite{BBmodel}, and the relevance model \cite{temporalEffects}, sequentially build off of each other and the foundational growth principal of preferential attachment.

\subsection{Barab\'asi-Albert Model}
The well known Barab\'asi-Albert model \cite{statMech} grows a network by adding one node with $m$ edges to the network at each time step. The probability $\Pi_i$ of attaching an edge to an existing node is dependent on that existing node's degree $k_i$. 

\begin{equation}
\Pi_i =\frac{k_i}{\sum\limits_{j} k_j}
\end{equation}

The network is grown in this fashion, adding one node at a time, until the desired size is achieved. This preferential attachment of adding edges to highly connected nodes will drive the network to a scale-free distribution of nodes with $k$ degrees.

\subsection{Bianconi-Barab\'asi Model}

One drawback of the Barab\'asi-Albert model is that older nodes have more opportunity to gain edges which makes them more likely to gain newer edges \cite{richer}. This disadvantages new nodes in their ability to catch up in number of degrees of older nodes. Yet in real world networks we often see that newer nodes are able to rapidly gain new edges due to characteristics that are not related to how many current connections they have. Bianconi and Barab\'asi introduced the concept of using a fitness value $\eta$ which describes a nodes ability to attract new edges regardless of its degree \cite{BBmodel}. The Bianconi-Barab\'asi model extends the Barab\'asi-Albert model by applying fitness value $\eta$ to its preferential attachment probability to obtain a new probability $\Pi_i$.

\begin{equation}
\Pi_i =\frac{\eta_i k_i}{\sum\limits_{j} \eta_j  k_j}
\end{equation}

The fitness value $\eta$ does not change over time and how well the Bianconi-Barab\'asi model still reflects a power law degree distribution will depend on how the fitness values are determined. For most distributions of fitness values the network should still maintain an approximate power law distribution \cite{decayRelevance}.

\subsection{Relevance Model}

Another phenomena observed in real networks is the decay of interest of nodes as a network expands. The relevance model \cite{temporalEffects} simulates this by adding a monotonically decreasing function $f_R (\tau)$ \cite{decayRelevance} which decays older node's attachment probability as time steps progress. As the time $\tau_i$ increases for a given node it is increasingly penalized on its probability of attracting new nodes. In this setup new nodes all start with no penalty.

\begin{equation}
\Pi_i =\frac{\eta_i k_i f_R (\tau_i) }{\sum\limits_{j} \eta_j  k_j f_R (\tau_j)}
\end{equation}

Like the fitness values, the decay function can be setup to achieve different output degree distributions \cite{temporalEffects} but for the purposes of this study the goal is to try to maintain a networks power law distribution as much as possible and so a constantly decreasing function was used.

\section{Methodology}

For this study all models were implemented in Python using the NetworkX package \footnote{https://networkx.org/}. The models were first grown using one of the three models mentioned above and then a separate model, discussed below, was used to evolve them. This study was begun with the expectation that each base model would need its own unique model that could evolve it, based on its unique extra parameters; fitness values for the Bianconi-Barab\'asi model and decay function for the relevance model. After much trial an error though, only one model was needed to evolve all three base models. 

Each model was grown to a node size of 10,000 with each node adding 3 edges upon being added to the network. This results in approximately 30,000 edges for a given experiment as duplicated edges are not accounted for and edges are bi-directional. The fitness values for the Bianconi-Barab\'asi model were drawn from a uniform sample and the decay function was a constant .001 at each time step. For this study the evolution of the network happens after it has been grown but there is nothing that prevents running them in conjunction. One could run one step of growth followed by one step of evolution and so on. 

For the evolution of a network I ran 300 time steps where approximately 1,000 edges are added and deleted at each time step, thereby preserving the approximate total edges. The mechanism for adding and deleted the edges are explained in depth in the model sections. Results are then displayed on a log-log plot with the power law exponent of 2 plotted for reference.

\subsection{Evolution Model}

The model that was eventually developed worked for all 3 base models and extended upon the preferential attachment method described by Barab\'asi and Albert to select which nodes would add and remove edges during each time step of evolution. The method in this model is based off of work previously done by Deijfen and Lindholm in \cite{prefDeletion} but the mechanisms in which nodes and edges are chosen are modified due to simulation results. There are two parts to the model, the growth of edges and the deletion. The growth of the edges is the simple part and simply follows the base growth model. The deletion of the edges takes an extra step, described in detail, in order to maintain the power law distribution.

At each time step $n$ nodes are selected to have an edge removed and $j$ nodes are selected to have an edge added. For this study $n = j = 1000$. For the $n$ nodes selected to grow an edge the target node for them to attach to is selected via the network's respective growth model approach; the Barab\'asi-Albert method, the Bianconi-Barab\'asi method, or the relevance method. This grows the model in accordance with preferential attachment and that network's growth model. To compliment this growth we need a way to delete edges. If we were to assume that evolving using the same mechanism as growing and perhaps delete edges randomly is sufficient, the network would quickly be dominated by high degree nodes as they start collecting all of the edges. Without new nodes to increase low degree node counts our network loses the target power law distribution of degrees.

\begin{figure}[ht]
    \centering
    \includegraphics[width=\linewidth]{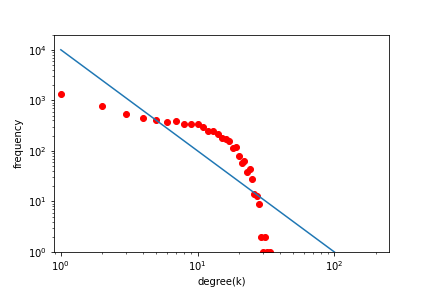}
    \caption{Evolution of a network where the edges selected to delete were based off of the  Barab\'asi-Albert model, degree distributions are approaching a normal distribution. Note the plot is in log-log format.}
    \label{fig:delete1}
\end{figure}

\begin{figure}[ht]
    \centering
    \includegraphics[width=\linewidth]{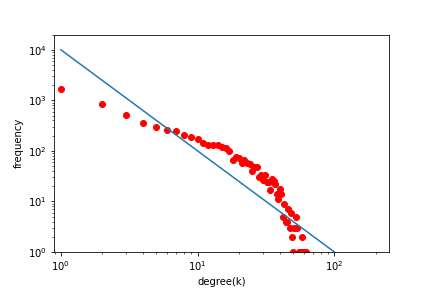}
    \caption{Evolution of a network where the edges selected to delete were based off of the inverse of the  Barab\'asi-Albert model, a closer approximation of the power law distribution}
    \label{fig:delete2}
\end{figure}

In order to maintain degree distributions we need to increase the number of low degree nodes while maintaining the low frequency of high degree nodes. So how do we do this through selecting edges to delete. If we look back to the growth step of this method we see that random nodes are selected to grow, and due to the preferential attachment a majority of our nodes are low degree nodes. This means that as the network evolves low degree nodes, on average, are selected more to grow than high degree nodes, at least in the beginning. Which means our selection for deletion of edges needs to bias towards low degree nodes, which might run counter-intuitive. This is because low degree nodes are steadily converging at a higher degree plateau as they are randomly selected to have an edge added. Yet we want lower degree nodes to have higher frequency than each higher degree nodes so we need to increasingly remove connections from nodes in the lower half of the degree distribution. One method to accomplish this is to reuse the Barab\'asi-Albert method, but now we don't want to heavily weight the high degree nodes, we want to do the opposite, see figures \ref{fig:delete1} and \ref{fig:delete2}.

Let $V_r$ be the $j$ nodes that have been selected to have an edge removed. The probability $\Pi_i^d$ of a given edge of node $ u \in V_r$ being removed is the inverse Barab\'asi-Albert probability of each node $w_i$ connected to $u$ forming edge $(u, w_i)$. The probability of $w_i$ being selected is now its degree ratio to all other opposing nodes connected to $u$, here labeled as $w_j$ instead of $k_j$ in previous equations. For the three base models, the probability $\Pi_i^d$ of a node having an edge deleted has been modified such that the probabilities are now inverted from the original model by subtracting from a maximum value based on the network. Once an opposing node $w_i$ is selected the edge between $(u,w_i)$ is deleted.

The method of inversion does not assume that the probabilities or fitness values add up to 1, they are instead calculated by subtracting the degree of the node from the maximum degree value of the network plus some $\beta$ value. This moves high degree values closer to zero and low degree values closer to the maximum degree that currently exists in the network. We add the $\beta$ value so that no node goes to zero and it still has a probability of having an edge deleted, which in this case would be if it had the maximum number of connecting edges currently found in the network. Without the $\beta$ value the node with the highest degree in the network would have no chance of having an edge deleted. The method to chose deletion of edges based on various growth models can be seen here; Barab\'asi-Albert model can be seen in equation \ref{eq:eBA}, the Bianconi-Barab\'asi model in equation \ref{eq:eBB}, and the relevance model in equation \ref{eq:eRE}.

\begin{equation}
\Pi_i^d =\frac{max(k_j) + \beta  - w_i}{\sum\limits_{j} max(k_j) +\beta -  w_j}
\label{eq:eBA}
\end{equation}

\begin{equation}
\Pi_i^d =\frac{max(k_j) + \beta  -\eta_i w_i}{\sum\limits_{j}max(k_j) + \beta  - \eta_j  w_j}
\label{eq:eBB}
\end{equation}

\begin{equation}
\Pi_i^d =\frac{max(k_j) + \beta -\eta_i w_i f_R (\tau_i) }{\sum\limits_{j} max(k_j) + \beta -\eta_j  w_j f_R (\tau_j)}
\label{eq:eRE}
\end{equation}
Earlier, from when I said that the evolution model was the same for all three growth models, this may seem like a violation of that axiom. But, in implementation, it really is the same model. If the decay function does nothing then the relevance model is the Bianconi-Barab\'asi model. If the fitness value is constant then the Bianconi-Barab\'asi model is the Barab\'asi-Albert model. Therefor we can use the same calculations for all three selections, we just need to use appropriate stand-ins for models with less parameters.

 The method described here makes edges to lower degree nodes more likely to be deleted which counters the impact of randomly selecting nodes to grow \footnote{Using the Barab\'asi-Albert model to select nodes to grow was also attempted, but led to a normal distribution (specifically the one seen in figure \ref{fig:delete1}) }. Edges are still likely to be added to high degree nodes but the growth low degree nodes see is kept in check by this inverted selection mechanism, again see figure \ref{fig:delete1} for when one doesn't invert the selection and figure \ref{fig:delete2} for when one does.

\section{Results}

\begin{figure*}[ht]
\centering
\begin{subfigure}{.3\textwidth}
  \centering
  \includegraphics[width=\linewidth]{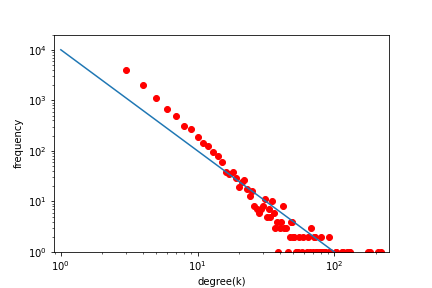}
  \caption{Barab\'asi-Albert Model\\ before evolution.}
  \label{fig:sfig1}
\end{subfigure}%
\begin{subfigure}{.3\textwidth}
  \centering
  \includegraphics[width=\linewidth]{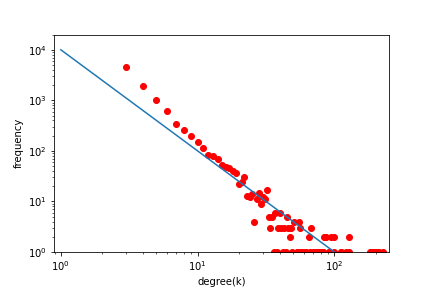}
  \caption{Bianconi-Barab\'asi Model\\ before evolution}
  \label{fig:sfig2}
\end{subfigure}
\begin{subfigure}{.3\textwidth}
  \centering
  \includegraphics[width=\linewidth]{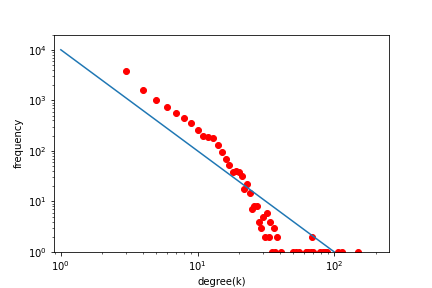}
  \caption{Relevance Model before\\ evolution}
  \label{fig:sfig3}
\end{subfigure}
\hfill{}
\begin{subfigure}{.3\textwidth}
  \centering
  \includegraphics[width=\linewidth]{nonfit_growth_loglog_last.png}
  \caption{Barab\'asi-Albert Model\\ after evolution.}
  \label{fig:sfig4}
\end{subfigure}
\begin{subfigure}{.3\textwidth}
  \centering
  \includegraphics[width=\linewidth]{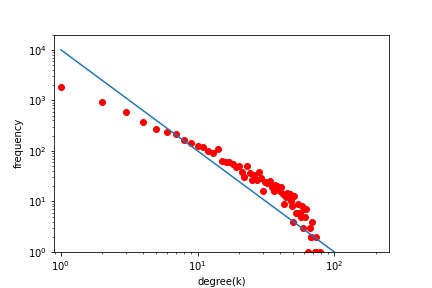}
  \caption{Bianconi-Barab\'asi Model\\ after evolution}
  \label{fig:sfig5}
\end{subfigure}
\begin{subfigure}{.3\textwidth}
  \centering
  \includegraphics[width=\linewidth]{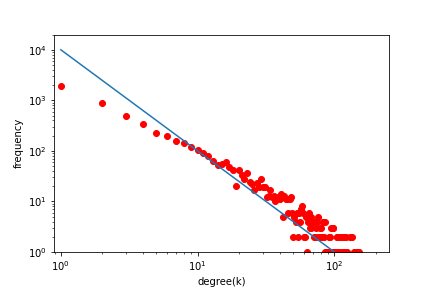}
  \caption{Relevance Model after\\ evolution}
  \label{fig:sfig6}
\end{subfigure}
\caption{Starting and end plots of evolved networks after 300 steps of evolution with approx. 1000 edges added and deleted at each step.}
\label{fig:results}
\end{figure*}

As can be seen in figure \ref{fig:results}, the evolution model had varying degrees of success with maintaining the power law degree distribution. Roughly speaking the more parameters that the model maintained, with the Barab\'asi -Albert model having none, the Bianconi-Barab\'asi model having one, and the relevance model having two, the better the model was at preserving the desired distribution.

As stated previously, the networks were grown to a node size of 10,000 before they were evolved, so results shown, both before and after evolution figures have 10,000 nodes and approximately 30,000 edges. In the top row of figure \ref{fig:results} one can see the state of each network after it was grown but before at has evolved. If we use subplot (a), the Barab\'asi-Albert model as the reference point, we can see that the Bianconi-Barab\'asi model produces a very similar initial distribution and that the relevance model shows a sharper decline in prevalence of higher degree nodes. This decline shows the counter impact the decay function has on old nodes gaining huge momentum in adding edges.

As for the evolution of the networks, the bottom row of figure \ref{fig:results}, we can see more dramatic changes. The Barab\'asi-Albert model is slowly transitioning to a more normal degree distribution\footnote{These experiments were run for only 300 steps, so longer simulations may show where these distribution eventually settle}. The Bianconi-Barab\'asi model looks as if it is maintaining a power law distribution better, but can we can still see evidence of the higher and lower degree nodes decreasing, a sign of transitioning to a normal distribution. The relevance model, however, appears to achieve an even more power law like distribution than its starting point as it evolves.

There is a caveat with the relevance model though. Since the relevance model adds a steady decay function to the node's probability, and this decay function is in use for the network's evolution, there is a limit to how long you could evolve the relevance  model. Even with a floor value, all nodes will eventually reach it and the decay function would serve no purpose. In practical usage the decay function may not be a constant decay and may vary per node as does the fitness value. Additionally, if the evolution was run intermittently with network growth, there would be new nodes with lower decay that routinely refresh the environment with more relevant nodes.

\section{Discussion}
This approach is obviously an approximation of real world dynamics, and misses a lot of the nuance of how networks evolve. There are a wide variety of network structures, techniques to generate them, and open problems to explore \cite{graphgen} of which this model only focused on techniques based on preferential attachment.  The eventual model described in this paper came about after a couple of false starts, described below in section \ref{negAttempts}, and has areas to grow, described in section \ref{nonPower} and section \ref{abm}.

\subsection{Negative Attempts}\label{negAttempts}
\begin{figure}
    \centering
    \includegraphics[width=\linewidth]{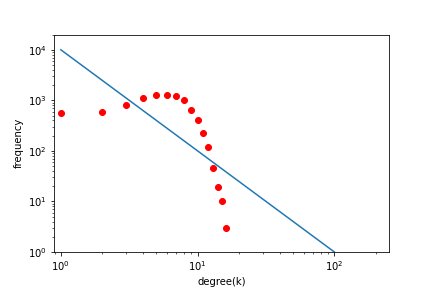}
    \caption{A network that has evolved to have a normal distribution of nodes with $k$ degrees.}
    \label{fig:bad1}
\end{figure}

This section will briefly go over some of the attempts that yielded negative results. Negative results occurred when the network's evolution led to a degree distribution that did not follow a power law, and more often than not, ended up similar to a normal distribution, see figure \ref{fig:bad1}. The original plan was to develop models that evolved the fitness value, introduced in the Bianconi-Barab\'asi model, in such a way that a node's fitness value would shift to maintain the overall power law distribution. 

Evolving a node's fitness value has similarities to the biological sciences where organisms fitness for survival evolves over time \cite{fitness}. I attempted various methods of evolving the fitness values ranging from adding a momentum parameter that increases and decreases the fitness value at each time step to scaling the fitness value by various functions (e.g. linear, log, and exponential). Each of these attempts ended with the network evolving to a distribution that was closer to normal than power law as seen in figure \ref{fig:bad1}.

There is still plenty of room to explore this mechanism as this reflects how real world networks evolve. Things grow and decay in popularity over time, reflecting a change in popularity. Where this my method may have come up short was in using a universal scaling function or where, when I did use a node specific momentum value, I did not evolve the momentum in a way that reflected realistic growth and decay.

\subsection{Non Power Law Models}\label{nonPower}

The focus of this paper has been in replicating power law distributions in networks. But not all networks exhibit this type of distribution \cite{erdosGraph, interdependentNetworks}. Can there be a unified model that, given different parameters, mimic different distributions? And what would those parameters look like? It has been shown that different parameters of the decay function can accomplish some of these tasks when growing a network \cite{temporalEffects}. Future work could explore a general function that can evolve networks and maintain desired distributions.

\subsection{Agent Based Models}\label{abm}
Agent based models are often times more complex to setup and run than the models described here, but they may be the better approach to simulate more complicated network dynamics. There has been work done using ABM's on simulating social networks \cite{synthSocial} as well as other types of networks that follow preferential attachment characteristics \cite{virusABM}. Simulation and analysis of ABM's that lead to scale-free networks can give insight into an individual actor's behavioral characteristics that will lead to these network structures. Additionally, modeling networks at the agent level can provide smaller granularity for time. Whereas time is marked, in the models shown here, at when a node or edge is added or deleted. For ABM's, there can be continuous measures of time where agents choose when or not to act, allowing for a wider range of behavior and outcomes.

\section{Conclusion}

What I have shown here is a two staged approach to growing and evolving networks. I have shown that it is possible to evolve networks that maintain their scale-free property while adhering to preferential attachment. I have approached this task from the simulation standpoint and adjusted models according to simulation results. I have not provided an analytic solution to these methods, which could be a possible future endeavor. There is also the possibility of creating a unified model that evolves and grows a model simultaneously by following simple preferential attachment principals. This type of model could be used as a baseline to compare more complicated methods emerging to synthesize networks.


\begin{thebibliography}{9}

\bibitem{statMech} 
Reka Albert, Albert-Laszlo Barabasi
\textit{Statistical mechanics of complex networks}
Reviews of Modern Physics, 2002

\bibitem{BBmodel} 
Ginestra Bianconi, Albert-Laszlo Barabasi
\textit{Competition and multiscaling in evolving networks}
Reviews of Modern Physics, 2002

\bibitem{prefDeletion} 
Mathias Lindholm, Maria Deijfen
\textit{Growing networks with preferential addition and deletion of edges}
Physica A: Statistical Mechanics and its Applications, 2009

\bibitem{decayRelevance} 
Jun Sun, Steffen Staab, Fariba Karima
\textit{Decay of Relevance in Exponentially Growing Networks}
Proceedings of the 10th ACM Conference on Web Science, 2018

\bibitem{temporalEffects} 
Matus Medo, Giuli Cimini, Stanislao Gualdi
\textit{Temporal effects in the growth of networks}
Physical Review Letters, 2011
 
 

\bibitem{richer} 
Faisal Nsour, Hiroki Sayama
\textit{Hot-Get-Richer Network Growth Model}
Complex Networks \& Their Applications IX. COMPLEX NETWORKS 2020. Studies in Computational Intelligence, vol 944. 

\bibitem{fitness} 
H. Allen Orr
\textit{Fitness and its role in evolutionary genetics}
Nat Rev Genet. 2009;10(8):531-539. doi:10.1038/nrg2603

\bibitem{graphgen}
Angela Bonifati, Irena Holubov\'a, Arnau Prat-P\'erez, Sherif Sakr
\textit{Graph Generators: State of the Art and Open Challenges}
ACM Computing Surveys April 2020 Article No.36

\bibitem{onlineSocial}
David Nettleton
\textit{A Synthetic Data Generator for Online Social Network Graphs}
Social Network Analysis and Mining 2016

\bibitem{erdosGraph}
P. Erd\"os and A. R\'enyi. 
\textit{On the evolution of random graphs.} 
Publ. Math. Inst. Hungar. Acad. Sci, 5:17–61, 1960.

\bibitem{interdependentNetworks}
Sergey Buldyrev, Roni Parshani, Gerald Paul, Eugene Stanley, Shlomo Havlin
\textit{Catastrophic cascade of failures in interdependent networks} 
Nature Volume 464, Issue 7291, 2010

\bibitem{smallWorldNetworks}
D. J. Watts, S.H. Strogatz 
\textit{Collective dynamics of 'small-world networks'} 
Nature Volume 393, Issue 6684, 1998

\bibitem{synthSocial}
Christopher Barrett, Richard Beckman, Maleq Khan, V.S. Anil Kumar, Madhav Marathe, Paula Tretz, Tridib Dutta, Bryan Lewis
\textit{Generation and Analysis of Large Sythetic Social Contact Networks} 
Proceedings of the 2009 Winter Simulation Conference

\bibitem{virusABM}
Muhammad Yasir, Muhammad Asif Habib, Muhammad Shahid, Mudassar Ahmad
\textit{Agent-based Modeling and Simulation of Virus on a Scale-Free Network} 
Proceedings of the International Conference on Future Networks and Distributed Systems 2017


\bibitem{socialSim}
Jim Blythe, Emilio Ferrara, Di Huang, Kristina Lerman, Goran Muri\'c, Anna Sapienza, Alexey Tregubov
\textit{The DARPA SocialSim Challenge: Massive Multi-Agent Simulations of the Github Ecosystem} 
Proceedings of the 18th International Conference on Autonomous Agents and MultiAgent Systems 2019
\end{thebibliography}
\end{document}